\documentclass[preprintnumbers, prd,onecolumn, floatfix, superscriptaddress, nofootinbib]{revtex4-2}

\usepackage{setspace}
\usepackage{graphicx}
\usepackage{subfig}
\usepackage{epsfig}
\usepackage{bm}
\usepackage{amssymb}
\usepackage{float}
\usepackage{amsmath}
\usepackage{dcolumn}
\usepackage[colorlinks]{hyperref}
\usepackage[usenames,dvipsnames]{color}
\usepackage{enumitem}
\usepackage[utf8]{inputenc}
\hypersetup{ breaklinks=true, pdfstartview={FitH}, colorlinks=true, linkcolor=blue, citecolor=red, filecolor=magenta, urlcolor=blue, anchorcolor=green, linktocpage=true }

\renewcommand{\(}{\left(}
\renewcommand{\)}{\right)}

\def\doi{http://doi.org}

\def\( \){$ $}

\newcommand{\be}{\begin{equation}}
\newcommand{\ee}{\end{equation}}
\newcommand{\bano}{\begin{eqnarray*}}
\newcommand{\eano}{\end{eqnarray*}}
\newcommand{\ba}{\begin{eqnarray}}
\newcommand{\ea}{\end{eqnarray}}
\newcommand{\no}{\nonumber}

\begin{document}

\title{Role of viscous fluid in FLRW model with observational constraints }
\author{Anirudh Pradhan}
\email{pradhan.anirudh@gmail.com (Corresonding Author)}
\affiliation{Centre for Cosmology, Astrophysics and Space Science (CCASS), GLA University, Mathura-281 406, Uttar Pradesh, India}
\author{G. K. Goswami}
\email{gk.goswami9@gmail.com}
\affiliation{Department of Mathematics, Netaji Subhas University of Technology, New Delhi-110 078, India}

\begin{abstract}
\begin{singlespace}
In this study, we investigate the role of bulk viscosity in the evolution of a spatially flat Friedmann-Lemaître-Robertson-Walker (FLRW) universe dominated by dust. By parameterizing the bulk viscosity as $\tilde{p}= -3(l + m (H'(t) + H^2) + n H^2) H$, we derived the modified Einstein field equations and reformulate them in terms of the redshift z. Using the Hubble datasets with 46 data points and Pantheon+ compilation with 1701 supernova measurements, we estimated the model parameters ($H_0$, l, m, n) through $\chi^{2}$ minimization and refined them using Markov  chain Monte Carlo (MCMC) simulations. Our analysis reveals a transition redshift $z_t = 0.585$, marking the universe's shift from decelerated to accelerated expansion, and a current deceleration parameter  $q_0 =- 0.705 $ consistent with the $\Lambda$CDM model. The current age of the universe obtained from  our model was $t_{0}$ = 14.5734 Gyrs. Additionally, we estimate the present Hubble constant, $H_0$ , to be 68 km/s/Mpc based on the Hubble datasets and approximately 73 km/s/Mpc using the Pantheon+ datasets. The best-fit cosmological parameters indicate a recent transition to accelerated expansion at $z_{t} \approx 0.55$, a present-day deceleration parameter $q_{0} \approx -0.55$, a Hubble constant in the range $67 -- 74 kms^{-1} Mpc^{-1}$, and a cosmic age $t_{0} \approx 13.7$ Gyr, consistent with current observational constraints. 
This disparity highlights the ongoing Hubble tension-a discrepancy between locally measured values of  $H_0$ (e.g., via supernovae) and values inferred from the early universe (e.g., CMB observations). Our findings are consistent with the larger trend of greater $H_0$ values from late-universe observations compared with early-universe predictions, highlighting the need for more research into the underlying physical or systematic sources of this tension. We further evaluate the evolution of the cosmological parameters, including the equation of state parameter $\omega(z)$, energy density $\rho(z)$ and pressure $p(z)$. Statefinder diagnostics $(r,s)$ demonstrate the versatility of the model, capturing its deviation from and convergence toward $\Lambda$CDM at specific epochs. This study emphasizes the importance of bulk viscosity in explaining the universe's accelerated expansion and provides a strong framework for assessing various cosmological theories.

\end{singlespace}

\end{abstract}

\maketitle


\section{Introduction}{\label{sec-1}}
Observational data \cite{SupernovaSearchTeam:1998fmf, SupernovaCosmologyProject:1998vns, Perlmutter:1999jt} have indicated an accelerating phase of the universe over the past few decades. The large-scale structure and cosmic microwave background radiation (CMBR) support this hypothesis. The Einstein equation is considered the standard equation, so modification of this Einstein equation is the path that leads to the goal of explaining this acceleration in the universe. One  approach is to consider specific forms for the energy-momentum tensor that modify the right-hand side of the Einstein equation, which has negative pressure and exerts the presence of exotic energy that is Dark Energy. The other is modifying the geometry of space-time, which modifies the left-hand side of Einstein's equation. These include a particular class related to modified theories of gravity. Various types of modified theories of gravity have been proposed to explain the current acceleration, such as $ f(R) $ gravity \cite{Starobinsky:2007hu, Capozziello:2008qc} where an arbitrary function depends on Ricci Scalar $ R $, $ f(R, T) $ gravity \cite{Harko:2011kv,Capozziello:2014bqa, Singh:2024ckh, Singh:2024kez} which depends on the Ricci Scalar $ R $ along with a trace of the energy-momentum tensor $ T $, Gauss-Bonnet theory \cite{Odintsov:2015uca, Wu:2015maa, Mustafa:2020jln}, symmetric teleparallel gravity ($ f(Q) $) gravity , and $f(Q, T) $ gravity in which the non-metricity tensor $ Q $ is coupled with a trace of the energy-momentum tensor $ T $, among others.\\
	
The expansion of the universe is the outcome of negative pressure in the cosmic fluid. This has led to  profound interest in studying cosmic fluids as a mixture of two components: a standard fluid component and a dark energy component \cite{Brevik:2012nt}. Dispersed processes such as heat transport, bulk viscosity, and shear viscosity play an important role in cosmic expansion. Among these, bulk viscosity has gained significant attention because of its potential impact on the early stages of cosmic evolution. The inclusion of bulk viscosity in the cosmological model may resolve singularities and explain the entropy generation. Several studies have examined the effects of bulk viscosity, including Misner \cite{Misner:1967uu}, who studied cosmological models with bulk viscosity and showed that  big-bang singularity can be avoided. Weinberg \cite{Weinberg:1971mx} derived general formulas for the bulk viscosity, shear viscosity, and heat transport in cosmology. Other studies \cite{Nightingale:1973, Heller:1975} have further explored the implications of viscosity in relativistic and isotropic cosmologies.\\

Given the success of non-viscous cosmic fluid models in the literature, it is natural to extend these models to include viscosity, particularly when considering turbulence effects. The addition of a viscosity term to the Friedmann equation raises the question of whether the original singularity in cosmology persists. Various viscosity theories have been proposed, starting from the first-order relativistic fluid theory by Eckart \cite{N1} and Landau-Lifshitz \cite{N2}, to the second-order relativistic theory introduced by Israel \cite{N3} and further developed by Israel and Stewart \cite{N4}. Despite their complexity, these approaches have helped understand the thermodynamic behavior of the Universe. Singh \cite{N5} considered the effect of bulk viscosity on the early evolution of universe for a spatially homogeneous and isotropic Robertson–Walker model. Recently, Zhadyranova et al. \cite{R1} studied an f(R,T) cosmological model that included the bulk viscosity. They considered two instances for the bulk viscosity coefficient: a time-dependent form $\xi = \xi_{0} + \xi_{1}H$ (Case I) and a constant viscosity (Case II). Using the Hubble $H(z)$ and Pantheon+SH0ES datasets, \cite{R1} used Bayesian approaches, specifically Markov chain Monte Carlo (MCMC) techniques implemented in the emcee Python package, to constrain model parameters. These findings show that the f(R,T) model with bulk viscosity provides a consistent and credible description of late-time cosmic acceleration, which is consistent with the observational data. Nandi and Choudhuri \cite{R2} proposed a feasible cosmological model with bulk viscosity and decaying vacuum density, interpreted as dark matter and dark energy in spatially FLRW spacetime. The presence of DM and DE necessitates interaction between these components. The authors \cite{R2} performed an MCMC analysis to limit the model parameters using the chi-square minimization method. They used two observational datasets: the CC dataset and Type Ia Supernova Pantheon dataset, with 31 and 1,048 data points, respectively. They also used the Pantheon+CC dataset. Villalobos et al. \cite{R3} recently investigated whether viscous cold dark matter ($\Lambda$CDM) in a $\Lambda$-dominated FLRW universe could ease the Hubble tension while fulfilling thermodynamic restrictions by studying both flat and curved geometries.  \\

In recent years, various studies have examined bulk viscosity using different modified theories of gravity.  Cardenas et al. \cite{Cardenas:2020exv} explored the cosmic expansion with matter creation and bulk viscosity. Solanki et al. \cite{Solanki:2021qni} investigated accelerated expansion with bulk viscosity in modified $ f(Q) $ gravity. Other studies \cite{ Gomez:2022qcu} have examined the behavior of dark energy with bulk viscosity in different cosmological models. The bulk viscosity coefficient of the form $\xi = \xi_{0} + \xi_{1} H + \xi_{2}\left(\frac{\dot{H}}{H } + H \right)$, was used by Gadbail et al. \cite{N11} to examine bulk viscous cosmological models in Weyl-type $f(Q,T)$ gravity.Dixit and Pradhan \cite{N12} recently explored the effects of bulk viscosity within the framework of  $f(T,B)$ gravity using a time-dependent viscosity model with a specific Hubble parameter expression, incorporating observational constraints. Similarly, Sasidharan and Mathew \cite{N13} examined a cosmological model dominated by bulk viscous matter, where the total bulk viscosity coefficient was given by $\xi = \xi_{0} + \xi_{1}\frac{\dot{a}}{a} + \xi_{2}\frac{\ddot{a}}{\dot{a}}$. A causal bulk viscous cosmological model with a bulk viscous coefficient proportional to the energy density of the cosmic fluid with variables $G$ and $\Lambda$ was studied by Belinchon et al. \cite{N14}. Additionally, Ren and Meng \cite{N15} investigated a cosmological model with a viscous medium (dark fluid), characterized by an effective equation of state with a time-dependent bulk viscosity of the above form of $\xi$. Furthermore, bulk viscosity in $F(T, T_{G})$ gravity was analyzed using a time-dependent viscosity model \cite{N16}. Several other studies have explored cosmological models that incorporate bulk viscosity under different scenarios \cite{N17,N18,N23,N24,N25,N26,N27}. Bulk viscous effects can act as an effective dark energy component by generating negative pressure in the cosmic fluid. In this sense, they can reproduce late-time accelerated expansion similar to that obtained in modified gravity frameworks. Therefore, the inclusion of these references provides a broader theoretical context for comparing different mechanisms that account for cosmic acceleration.
 \\

In this study, we investigated a spatially homogeneous and isotropic Friedmann-Lemaître-Robertson-Walker (FLRW) universe filled with a bulk viscous fluid. Bulk viscosity modifies the effective pressure of the cosmic fluid by introducing a negative contribution, which can mimic dark energy-like behavior. We adopted a generalized parameterization of the bulk viscosity coefficient $\xi$, expressed as a function of the Hubble parameter $H$ and its time derivative $\dot{H}$, follows
\[
\xi = l + m \frac{\ddot{a}(t)}{a(t)} + n \left(\frac{\dot{a}(t)}{a(t)}\right)^2,
\]
which, expressed in terms of the Hubble parameter $H = \frac{\dot{a}}{a}$, reads
\[
\xi = l + m \big(\dot{H} + H^2 \big) + n H^2.
\]
where $l$, $m$, and $n$ are constant parameters determined from the observational data.
This form was motivated by several considerations:

\begin{itemize}
	\item \textbf{Physical grounds:} The coefficient includes a constant term $1$, a term proportional to the acceleration of the scale factor $\frac{\ddot{a}}{a}$ (or equivalently $\dot{H} + H^2$), and a term proportional to the square of the Hubble parameter $H^2$. These correspond to the viscosity contributions related to the expansion rate and its change, capturing both velocity- and acceleration-dependent dissipative effects in the cosmic fluid. 
	
	\item \textbf{Generality and prior use:} Similar parameterizations have been employed in the literature \cite{N13, Ren:2010ckh, Belinchón:2004} to explore the bulk viscous effects in cosmology. Our choice is a natural extension that allows a flexible description of viscous behavior while remaining tractable analytically and numerically.
	
	\item \textbf{Observational motivation:} The inclusion of acceleration-dependent and velocity-dependent terms enables the model to capture richer phenomenology and better fit observational data, such as Hubble parameter measurements and supernovae luminosity distances, which were tested through parameter estimation in our analysis.
	
	\item \textbf{Alternative forms:} While other forms of $\xi$ are possible (e.g., depending only on $H$ or on the energy density $\rho$), the chosen form balances the  complexity and physical insight, allowing us to investigate the combined effects of constant, velocity-, and acceleration-dependent viscosity within a unified framework.
\end{itemize}
Thus, the formulation extends previous models that consider bulk viscosity proportional to $H$ or $H^2$ alone, allowing for a richer description of the viscous effects on cosmic evolution. Our goal is to explore whether such viscous dynamics, without invoking modifications to General Relativity or exotic dark-energy components, can provide a viable explanation for the late-time acceleration of the universe.
To this end, we solved the Einstein field equations coupled with this viscous fluid model and constrained the parameters $l$, $m$, and $n$ using observational datasets including Hubble parameter measurements and the Pantheon+ supernova compilation. This approach aims to clarify the role of bulk viscosity in the cosmic evolution within the standard gravitational framework. \\

We aimed to analyze the impact of bulk viscosity on key cosmological parameters, including the deceleration parameter, equation of state parameter, energy density, and pressure. Additionally, we sought to constrain the bulk viscosity parameters using observational datasets such as Hubble and Pantheon data.\\
The objectives of this study are as follows:
\begin{itemize}
\item To formulate the field equations incorporating bulk viscosity in the framework of modified gravity.
\item  The unknown model parameters were estimated  using observational datasets, including the Hubble and Pantheon datasets.
\item To analyze the effects of bulk viscosity on cosmological parameters and compare different models.
\item To conduct a state-finder analysis to distinguish the proposed model from standard dark energy models.
\item To investigate the age of the Universe by converting redshift into cosmic time.
\item To summarize key findings and discuss their implications in the broader context of cosmology.
\end{itemize}

We also know that flexibility in suggesting various forms of bulk viscosity reduces the prediction potential of such models, lowering their scientific importance. However, with further exploration, we can enter a new era of bulk viscosity research in cosmology. Despite its challenges (for example, the arbitrariness of the viscosity function), viscous cosmology remains a promising area of theoretical inquiry with improved micro-physical foundations and empirical constraints. The idea that bulk viscosity could drive late-time accelerated expansion of the universe is based on modifying the effective pressure in the cosmic fluid. In standard cosmology, the universe is modeled as a perfect fluid. However, real fluids can have dissipative processes, such as bulk viscosity, which arise in systems out of equilibrium and affects the expansion dynamics. Bulk viscosity modifies the effective pressure of the cosmic fluid,
	$ p_{eff} = p -\xi H $, where $p$ is the standard pressure, $\xi$ is the bulk viscosity coefficient, and $H$ is the Hubble parameter. If $\xi > 0$, this results in an additional negative pressure, which can mimic dark energy-like behavior and potentially lead to accelerated expansion. The negative effective pressure can dominate the matter energy density at late times. This leads to a repulsive gravitational effect, similar to a cosmological constant. Hence, the bulk viscosity can act as a dark energy alternative or complement. Depending on the form of $\xi(H)$, various expansion scenarios are possible, including de Sitter-like acceleration, phantom behavior, and avoidance of singularities. The following are important references that detail the development and implications of viscous cosmology \cite{G1,G2,G3,G4,G5}. Thus, the study of bulk viscosity in cosmology is significant because it provides a scientifically motivated extension to standard cosmological models, allowing researchers to investigate dissipative processes in the evolution of cosmos. \\
	
The remainder of this paper is organized as follows. Metric and field equations are introduced in Sec. II. In Section III, we use a variety of datasets, including the Hubble and Pantheon datasets, to estimate the unknown parameters. Error bar plots were then used to compare our computed models. In Sec. IV, we discuss the implications of the  bulk viscosity on different cosmological parameters. The state-finder analysis is explained in Section V. Section VI discusses the conversion of redshift into time and the age of the universe. Finally, we summarize our findings and provide conclusions in Section VII.
\section{Metric and Field Equations}
We consider a spatially isotropic and homogeneous Universe FLRW metric
	\begin{equation}\label{1}
		ds^2 = -dt^2 + a^2(t) (dx^2 + dy^2 + dz^2),
	\end{equation}
where $ a(t) $ is the scale factor.
	The stress energy-momentum tensor in the presence of bulk viscosity is given by 
	\begin{equation}\label{2}
		T_{ij} = (\tilde{p} + \rho) u_i u_j + \tilde{p} g_{ij},
	\end{equation}
	where,~~
    \begin{equation}\label{2a}
\tilde{p}= p_m - 3 \xi H,
	\end{equation}
	  $p_m$ is the matter pressure and  $H = \frac{\dot{a}(t)}{a(t)}$ is the Hubble parameter. The over head dot $(^{.})$ represents differentiation with respect to `t' here and elsewhere. In Eq. (3), $- 3 \xi H $ is the pressure due to bulk viscosity. It is negative, so it helps in producing acceleration in the universe which in turn may play the role of dark energy in the universe. In the following analysis, we denote the effective pressure $\tilde{p}$ simply by $p$ for notational convenience.\\
	  
	   In this study, we consider a parameterized bulk viscosity of the form 
	\begin{equation}\label{3}
		\xi = l + m \frac{\ddot{a}(t)}{a(t)} + n \bigg( \frac{\dot{a}(t)}{a(t)} \bigg)^2,
	\end{equation} 
where $l$, $m$ and $n$ are constant parameters. Eq. (4) has a clear  theoretical and observational motivation, particularly in modern cosmology, where effective fluid descriptions are commonly used. Equation (4) is a simple but physically motivated extension that includes intrinsic dissipation, acceleration-driven non-equilibrium effects, and expansion-rate-dependent adjustments. This form accurately represents both early- and late-time cosmic dynamics and can simulate modified gravity contributions while being thermodynamically consistent. Parameters $l$, $m$, $n$ can be constrained using Pantheon+(SNe Ia), BAO, Cosmic chronometers, and CMB shift parameters.  
In terms  of the Hubble parameter $H$, the above equation can be written as 
	\begin{equation}\label{4}
		\xi = l + m (\dot{H}(t) + H^2(t)) + n H^2(t).
	\end{equation}
The form of the viscous fluid in our formulation incorporates a parameterized bulk viscosity $\xi$ which depends on both the Hubble parameter $H$ and its  derivative $\dot{H}.$ Let us analyze its implications:\\
	1. Role of Bulk Viscosity in Cosmic Acceleration:
\begin{itemize}
	\item The effective pressure $\tilde{p}$ in Eq. (3) is modified by the term $- 3 \xi H$, which represents the contribution of the bulk viscosity.
	\item Because $ \xi H$   is generally positive, this results in a more negative  $\tilde{p}$, which helps drive cosmic acceleration, mimicking the behavior of dark energy.
	\end{itemize}
	2. Functional Form of Viscosity $\xi$:
	\begin{itemize}
		\item The viscosity function in Eq. (5) is parameterized as $\xi = l + m (\dot{H}(t) + H^2(t)) + n H^2(t).$
		\item A constant term (l), which represents a baseline viscosity.
		\item The term is proportional to the acceleration of expansion ($\dot{H}(t)$), meaning viscosity is influenced by changes in the Hubble rate.
		\item A term proportional to the square of the expansion rate ($H^2(t)$), linking the viscosity to the expansion speed itself.
		\end{itemize}
		3. Physical Interpretation of the Parameters:
		\begin{itemize}
			\item  A constant bulk viscosity that exists even when expansion is negligible.
			\item  Controls how viscosity responds to the acceleration of expansion, meaning that it can influence the transitions between the deceleration and acceleration phases.
			\item  Determine how viscosity scales with the square of the Hubble parameter, affecting early-universe dynamics when $H$ is large.
		\end{itemize}
			4. Connection to Dark Energy and Modified Gravity:
			\begin{itemize}
				\item Because the bulk viscosity acts as an effective pressure, its negative contribution can lead to accelerated expansion, similar to a dark energy component.
				\item The specific form of $\xi$ suggests a connection to modified gravity models( e.g. $f(Q,T)$ or $f(T,B)$ gravity), where additional  terms arise naturally.
				\end{itemize}
			5. Comparison with Other Viscosity Models:
			\begin{itemize}
				\item This form of {\color{red}$\xi$} generalizes earlier models, such as:\\
				(a) $\xi \propto H$(often used in simple bulk viscous cosmology).\\
				(b)  $\xi \propto H^2$(motivated by thermodynamic and kinetic considerations).
			\end{itemize}
	The Einstein field equations (EFE) of general relativity are described as
	\begin{equation}\label{5}
			R_{ij}- \frac{1}{2}g_{ij} = \frac{8\pi G}{c^4} T_{ij}
	\end{equation}
	The EFE corresponding to the metric in Eq.(\ref{1}) and The stress energy-momentum tensor   Eq. (\ref{2}), are obtained as:
	\begin{equation*}
	3 H(t)^2 = \frac{8 \pi G}{c^4} \rho
	\end{equation*}
	\begin{equation*}
	2 \dot{H}(t) +3 H(t)^2 = - \frac{8 \pi G}{c^4} \tilde{p}.
	\end{equation*}
	 We assumed that $ \frac{8 \pi G}{c^4} \equiv 1.$ In the viscous fluid model at the present  dust-filled universe,  matter pressure  is considered as $ p_m =0,$ the
	 simplified EFE are obtained as follows:
	\begin{equation}\label{6}
	3 H(t)^2 = \rho
	\end{equation}
	\begin{equation}\label{7}
	2\dot{H}(t) +3 H(t)^2 =  3(l + m (\dot{H}(t) + H(t)^2) + n H(t)^2) H(t) 
	\end{equation}
	To evaluate unknown model parameters $l$, $m$, and $n$, we  used observational data sets such as the Hubble  and Pantheon+ Datasets. For this, we  need to express and solve the Einstein field equations  in terms of the redshift $z$ instead of time. We will use the following transformation equations for the scale factor $a(t)$ and $z$.
	\begin{equation*}
		\frac{a_0}{a(t)} =1+z, \; \dot{H}(t) = -(1+z) H(z)  H'(z),
	\end{equation*}
   where $H'(z)$ is derivative of $H$ with respect to $z$. \\
	 Using these values, we obtain the replacement of Eqs. (7) and (8) as 
	\begin{equation}\label{8}
		3 H^2(z) =  \rho(z),
	\end{equation}
	and
	\begin{equation}\label{9}
		-2 (1+z) H(z) H'(z) +3 H^2(z) = 3(l + m (-(1+z) H(z) H'(z) + H^2(z)) + n H^2(z)) H(z) ,
	\end{equation}

	The simplified form of Eq. (\ref{9}) can be written as,
	\begin{equation}\label{10}
		3(l+ H(z) (-1+(m+n) H(z)))+ (2-3m H(z))(1+z) H'(z)=0.
	\end{equation}
Equation (10) was obtained by expanding Eq. (9), collecting derivative and algebraic terms separately, and dividing by the Hubble parameter $H(z)$. This form clearly separates the contribution of the expansion rate from its redshift evolution, making it suitable for analytical and numerical reconstruction of $H(z)$. 
 We numerically solved Eqs. (\ref{10}) for the Hubble parameter $H(z)$ using the initial condition $H(0) = H_0$. Because this equation contains the free parameters $l$, $m$, and $n$, it is integrated repeatedly for various parameter sets during observational analysis. In particular, these solutions were used in conjunction with Eqs. (\ref{chi}) to compute the $\chi^{2}$, allowing for direct comparison with the observational data in the subsequent sections. Furthermore, Eqs. (15) and (16) relate the distance modulus $\mu$ to the luminosity distance $D_L(z)$, which is  used in the observational analysis for comparison with supernova data, and depends on the numerical solution of $H(z).$
\section{Observational Analysis:}
\subsection{Hubble datasets}
 We consider the  Hubble parameter Table \cite{Bhardwaj} consisting of
46 data set of observed values of $H$ for various
redshifts in the range $ 0 \leq z \leq 2.36 $ with possible error in observations.
We used this data set to estimate the model parameters: $H_{0}$, $l$, $m$, and $n$ in multiple ways. First, we fitted the Hubble parameter function given by Eq. (\ref{10}) to the data methods with the least $\chi^{2}$.
Then, we can carry out the Markov chain Monte Carlo (MCMC) simulations method to further refine the estimations by assuming $\chi^{2}$ estimations as
initial guess. Recall that $\chi^{2}$ formula for the Hubble-function is as follows:
\begin{equation} \label{chi}
   {\chi^{2}}_{OHD}{(H_0,  l, m, n,)}=\sum\limits_{i=1}^{46}\frac{[H_{th}(H_0, l, m, n, z_{i})-H_{obs}(z_{i})]^{2}}{ \sigma^2 (z_{i})},
\end{equation}
where $ H_{th} $ and $ H_{obs} $ represent the theoretical and observational values of the Hubble parameter respectively. 
We now present the following figures 1 to show our work. 
\begin{figure}[H]\label{fig1}
	\centering
	a.\includegraphics[width=8cm,height=8cm,angle=0]{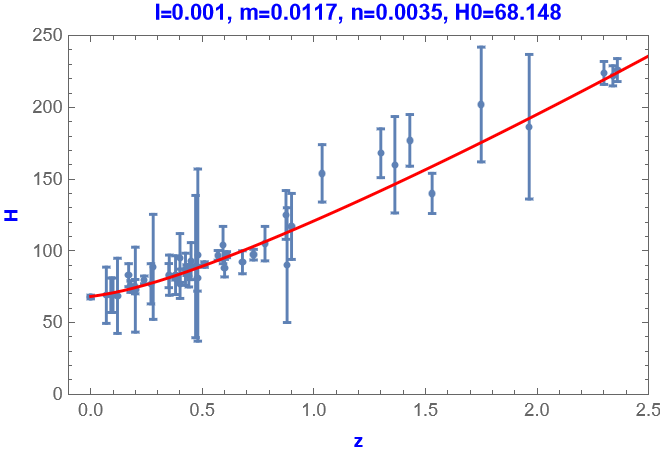}
	b.\includegraphics[width=8cm,height=8cm,angle=0]{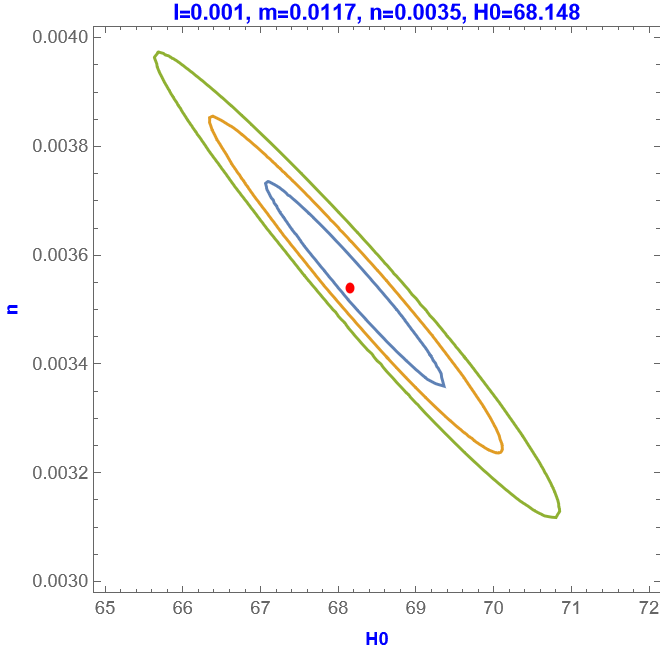}\\
 c.\includegraphics[width=15cm,height=15cm,angle=0]{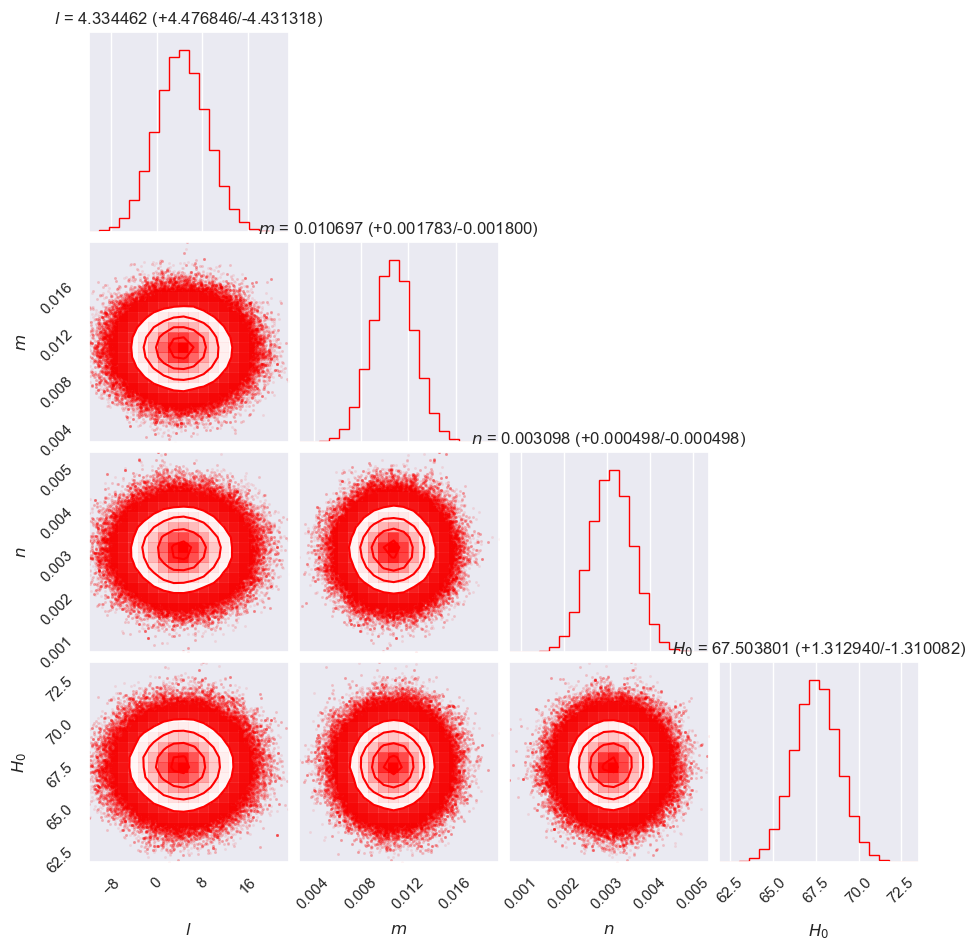}
\caption{Fig 1a: Error bars plot for Hubble parameter. Parameters estimation based on minimum $\chi^{2}$.	Fig 1b: Contour region plots for min $\chi^{2}$ plus 1$\sigma$, 2$\sigma$ and 3 $\sigma$.
	Fig 1c: MCMC simulation-based estimations of parameters. Estimated parameter values are described in the figures. }
\end{figure}


The figures show a comparison between theoretical predictions and observed data for the Hubble parameter (Fig 1a) and an MCMC corner plot for parameter estimation (Fig 1c). The theoretical curve (red line) fits the observed data (blue error bars) reasonably well, shows a good agreement. The plot clearly represents the error bars for the observed data, enhancing the visualization of the uncertainties. The corner plot effectively shows the posterior distributions of parameters and their correlations. The contour levels are clear, and the histograms at the top provide insights into the parameter uncertainties. The figures suggest a solid analysis with a good theoretical fit and a well-constrained MCMC parameter estimation. Fig 1(b) shows the  contour plot of the parameter pair $(H_0, n)$ showing $1\sigma,2\sigma$ and $3\sigma$ regions.
The estimations are presented in the following Table-1
\begin{table}[H]
	\centering
	\begin{tabular}{|c|c|c|}
		\hline
		Parameters & Least $\chi^2$ & MCMC simulation \\
		\hline
		$l$  & $0.001$  & $4.3344^{+4.4768}_{-4.4313}$ \\
		\hline
		$m$  & $0.0117$ & $0.0107^{+0.0018}_{-0.0018}$ \\
		\hline
		$n$  & $0.0035$ & $0.0031 \pm 0.0005$ \\
		\hline
		$H_{0}$ & $68.148$ & $67.504^{+1.131}_{-1.130}$ \\
		\hline
	\end{tabular}
	\caption{The best-fit values of the model parameters $l$, $m$, $n$, and $H_{0}$ for the best-fit $H(z)$ Hubble curve.}
	\label{table1}
\end{table}
\subsection{Pantheon Plus Datasets}
In this subsection, we use the Pantheon+ datasets compilation\cite{Pan2022} consisting of 1701 supernova datasets of red shift versus distance modulus to estimate the values for unknown constants $l$, $m$, $n$, and $H_{0}$. For this, we first fitted the distance modulus ($\mu$) function given by Eq. (\ref{eq15}) and (\ref{eq16}) to the observational distance modulus ( $\mu$) data using the method of least $\chi^2$ and estimates the model parameters. Finally, we perform Monte-Carlo simulations using the MCMC method to further refine the estimations by assuming $\chi^2$ estimations as the initial guess. 
We  used the following $\chi^{2}$ function  to estimate the values for unknown constants,
\begin{equation}\label{eq14}
 {\chi^2}_{Pan}(l, m, n, H0) = \sum\limits_{i=1}^{ 1701}\frac{[\mu_{th}(H_0, l, m, n, z_{i})-\mu_{obs}(z_{i})]^{2}}{err_{\mu} {(z_{i})}^{2}},
\end{equation}
where $ \mu_{th} $ and $ \mu_{obs} $ represent the  theoretical and observed distance moduli of the model respectively. $ err_{\mu}$ denoted the standard error. 
The distance modulus $ \mu(z) $ can be defined as 
\begin{equation}\label{eq15}
      \mu(z) = m- M = 5 \log D_l(z) + \mu_0,
\end{equation}
where $ m $ and $ M $ represent the apparent and absolute magnitudes of the standard source, respectively.
The Luminosity distance is defined as \cite{Copeland:2006wr},
\begin{equation}\label{eq16}
   D_l(z) = c (1+z)  \int_{0}^{z} \frac{dz*}{H(z*)},
\end{equation}
 and nuisance parameter $ \mu_0 = 25+ 5 \log(\frac{{H_0}^{-1}}{1Mpc} ) $.
 We present the following plots to illustrate the results.
\begin{figure}[H]\label{fig1}
	\centering
	a.\includegraphics[width=8cm,height=8cm,angle=0]{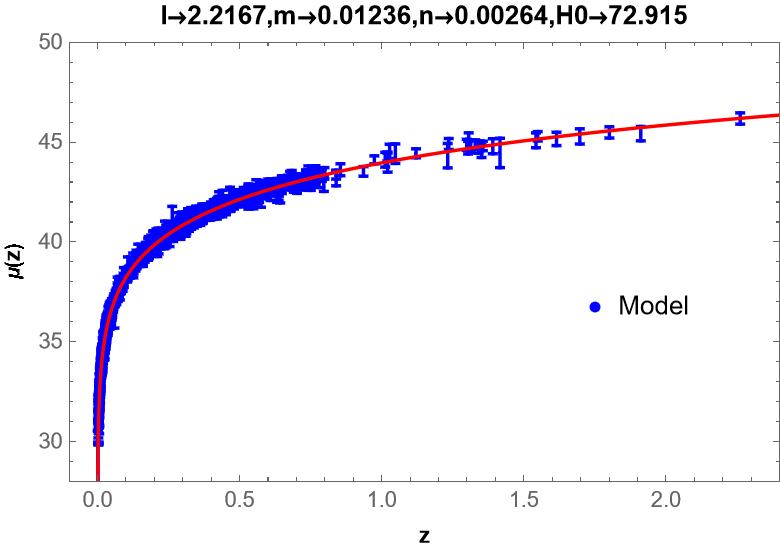}
	b.\includegraphics[width=8cm,height=8cm,angle=0]{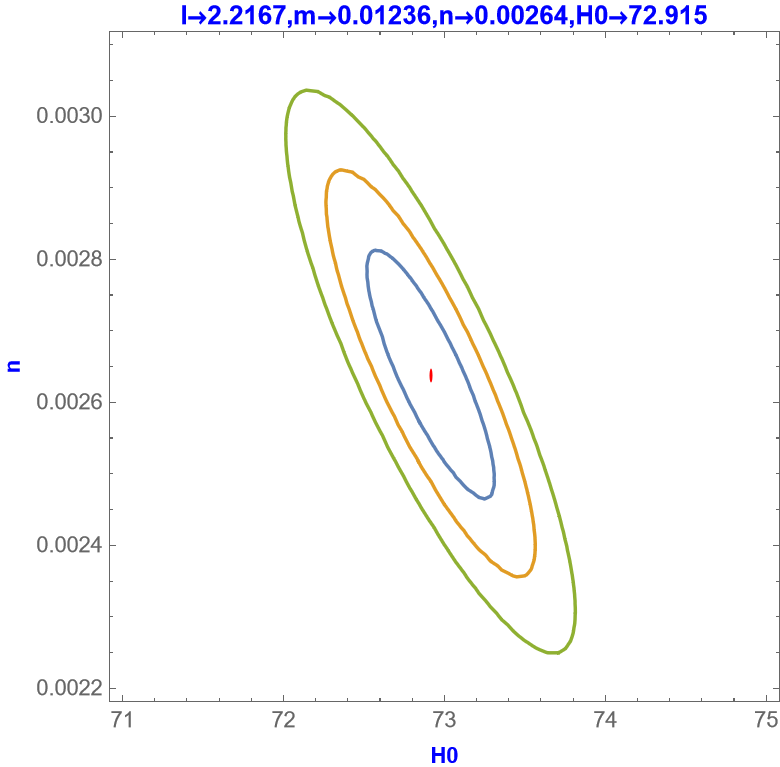}
	c.\includegraphics[width=15cm,height=15cm,angle=0]{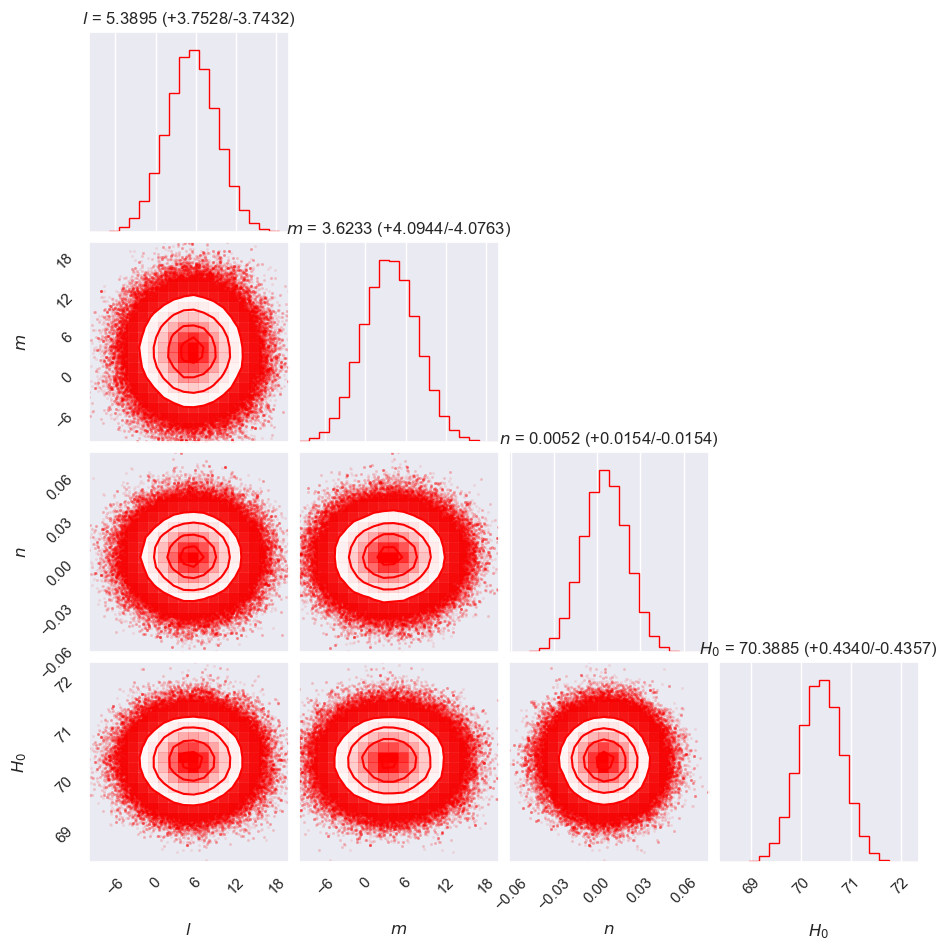}
	\label{fig2}
\end{figure}
 
 \begin{figure}[H]\label{fig1}
	\centering
		d.\includegraphics[width=10cm,height=10cm,angle=0]{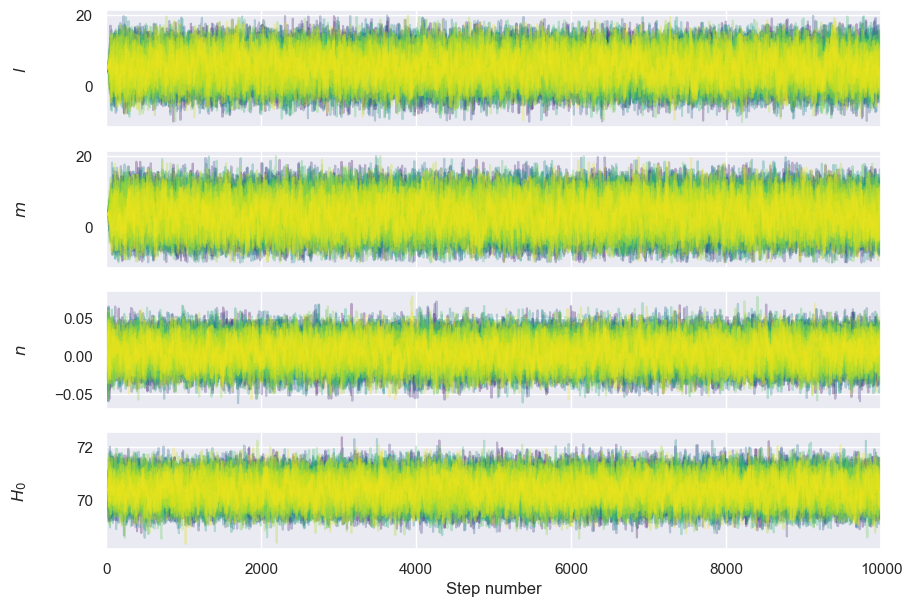}
	\caption{Fig 2a: Error bars plot for Distance modulus. Parameters estimation based on minimum $\chi^{2}$.
		Fig 2b: Contour region plots for min $\chi^{2}$ plus 1$\sigma$, 2$\sigma$ and 3 $\sigma$.
		Fig 2c and Fig 2d: MCMC simulation-based estimations of parameters and corresponding step number plots. Estimated parameter values are described in the figures. We have used  Pantheon+ data set for estimation }
	\label{fig2}
\end{figure}
The fig(2a) shows a comparison between the theoretical predictions and the observed data for the distance modulus. The theoretical curve (red line) fits the observed data (blue error bars) reasonably well, and shows good agreement. The plot clearly represents the error bars for the observed data, enhancing the visualization of the uncertainties. The corner plot in Fig 2(c)  shows the posterior distributions of the parameters and their correlations. The contour levels are clear, and the histograms at the top provide insights into the parameter uncertainties. The figures suggest a solid analysis with a good theoretical fit and  well-constrained MCMC parameter estimation. We have also presented a contour plot in Fig (2b) for min $\chi^{2}$ plus 1$\sigma$, 2$\sigma$ and 3$\sigma$. Fig (2d) is the step number plot for MCMC parameter estimation.
Our estimations are shown in the following Table-2

\begin{table}[H]
	\centering
	\begin{tabular}{|c|c|c|}
		\hline
		Parameters & Least $\chi^2$ & MCMC simulation \\
		\hline
		$l$  & $2.2167$  & $5.3895^{+3.7528}_{-3.7432}$ \\
		\hline
		$m$  & $0.01236$ & $3.6233^{+4.0944}_{-4.0763}$ \\
		\hline
		$n$  & $0.00264$ & $0.0052 {+0.0154}_{-0.0154}$ \\
		\hline
		$H_{0}$ & $72.915$ & $70.3885^{+0.4340}_{-0.4357}$ \\
		\hline
	\end{tabular}
	\caption{The best-fit values of the model parameters $l$, $m$, $n$, and $H_{0}$ for the best-fit $\mu(z)$ distance modulus curve.}
	\label{table2}
\end{table}
 Because the parameter estimations obtained from the two different observational datasets — the 46-point Hubble cosmic chronometer data and the 1701-point Pantheon+ distance modulus data, — show noticeable differences, it is useful to provide a comparative summary. To highlight these deviations in the estimated parameter values, we present Table 3, which consolidates the results from both datasets for direct comparison.
	\begin{table}[H]
	\centering
	\begin{tabular}{|c|c|c|}
	   \hline
		Parameters & MCMC simulation  & MCMC simulation  \\
		           & Hubble           &  Distance modulus \\
		\hline
		$l$   & $4.3344^{+4.4768}_{-4.4313}$ & $5.3895^{+3.7528}_{-3.7432}$ \\
		\hline
		$m$  & $0.0107^{+0.0018}_{-0.0018}$ &  $3.6233^{+4.0944}_{-4.0763}$  \\
		\hline
		$n$  & $0.0031 \pm 0.0005$ &  $0.0052 {+0.0154}_{-0.0154}$          \\
		\hline
		$H_{0}$ & $67.504^{+1.131}_{-1.130}$ &  $70.3885^{+0.4340}_{-0.4357}$   \\
		\hline 
	\end{tabular}
	\caption{The best-fit values of the model parameters $l$, $m$, $n$, and $H_{0}$ for the best-fit $H(z)$ Hubble and $\mu(z)$ distance modulus curves.}
	\label{table3}
\end{table}
\subsection{Union3 Supernova Likelihood}
\subsubsection{Type Ia Supernova Constraints}

Type Ia supernovae (SNe Ia) provide one of the most powerful probes of the late–time expansion history of the Universe. In this subsection, we use the binned Union3 supernova compilation, which contains measurements of the apparent magnitude $m_B$ at different redshifts.

For a given cosmological model, the theoretical prediction for the apparent magnitude is given by
\begin{equation}
	m_B^{\mathrm{th}}(z) = \mathcal{M} + 5 \log_{10}\left(D_L(z)\right),
\end{equation}
where $\mathcal{M}$ is a nuisance parameter that absorbs the absolute magnitude of the supernovae and other normalization constants. The luminosity distance $D_L(z)$ is computed from the cosmological model under consideration.

In our model, the luminosity distance is obtained by solving the coupled differential equations
\begin{align}
	\frac{dH}{dz} &= \frac{3\left(l + H(-1 + (m+n)H)\right)}{(1+z)\left(-2 + 3mH\right)}, \\
	\frac{dD_L}{dz} &= \frac{c(1+z)}{H(z)} + \frac{D_L}{1+z},
\end{align}
with initial conditions
\begin{equation}
	H(0) = H_0, \qquad D_L(0) = 0.
\end{equation}

Here $H(z)$ is the Hubble parameter, $H_0$ is the present Hubble constant, and $l$, $m$, and $n$ are free model parameters.
\subsubsection{Covariance Matrix and $\chi^2$ Estimator}

To compare theoretical predictions with the observational data, we construct the chi-square statistic using the full covariance matrix provided with the Union3 compilation. The $\chi^2$ function is defined as

\begin{equation}
	\chi^2_{\mathrm{SN}} =
	\Delta \mathbf{m}^T
	\mathbf{C}^{-1}
	\Delta \mathbf{m},
\end{equation}

where

\begin{equation}
	\Delta \mathbf{m} = \mathbf{m}_B^{\mathrm{obs}} - \mathbf{m}_B^{\mathrm{th}}
\end{equation}

is the vector of differences between the observed and theoretical apparent magnitudes. The matrix $\mathbf{C}$ represents the covariance matrix of the supernova data, which accounts for both statistical and systematic uncertainties.

The likelihood function is then given by

\begin{equation}
	\mathcal{L}_{\mathrm{SN}} \propto
	\exp\left(-\frac{\chi^2_{\mathrm{SN}}}{2}\right).
\end{equation}

The cosmological parameters are constrained by minimizing the $\chi^2$ function and by exploring the posterior distribution using the Markov Chain Monte Carlo (MCMC) technique.
\subsubsection{Parameter Estimation}

We estimate the cosmological parameters by performing a Markov Chain Monte Carlo (MCMC) analysis using the \texttt{emcee} ensemble sampler. The parameter space explored in this analysis consists of the set

\begin{equation}
	\Theta = \{l, m, n, H_0, \mathcal{M}\}.
\end{equation}

Uniform priors are adopted for all parameters within physically motivated ranges. The posterior distributions are obtained by sampling the likelihood function, and the final constraints are quoted as the median values with $1\sigma$ uncertainties corresponding to the 16th and 84th percentiles of the marginalized distributions.
\subsubsection{Results}

Using the Union3 supernova data, we obtain the following constraints on the model parameters:

\begin{align}
	l &= -70.77^{+30.87}_{-20.51}, \\
	m &= 0.0595^{+0.0241}_{-0.0207}, \\
	n &= 0.00536^{+0.00578}_{-0.00371}, \\
	H_0 &= 63.22^{+14.25}_{-9.42}, \\
	\mathcal{M} &= 24.72^{+0.45}_{-0.35}.
\end{align}

\begin{figure}
	\centering
	\includegraphics[width=0.50\textwidth]{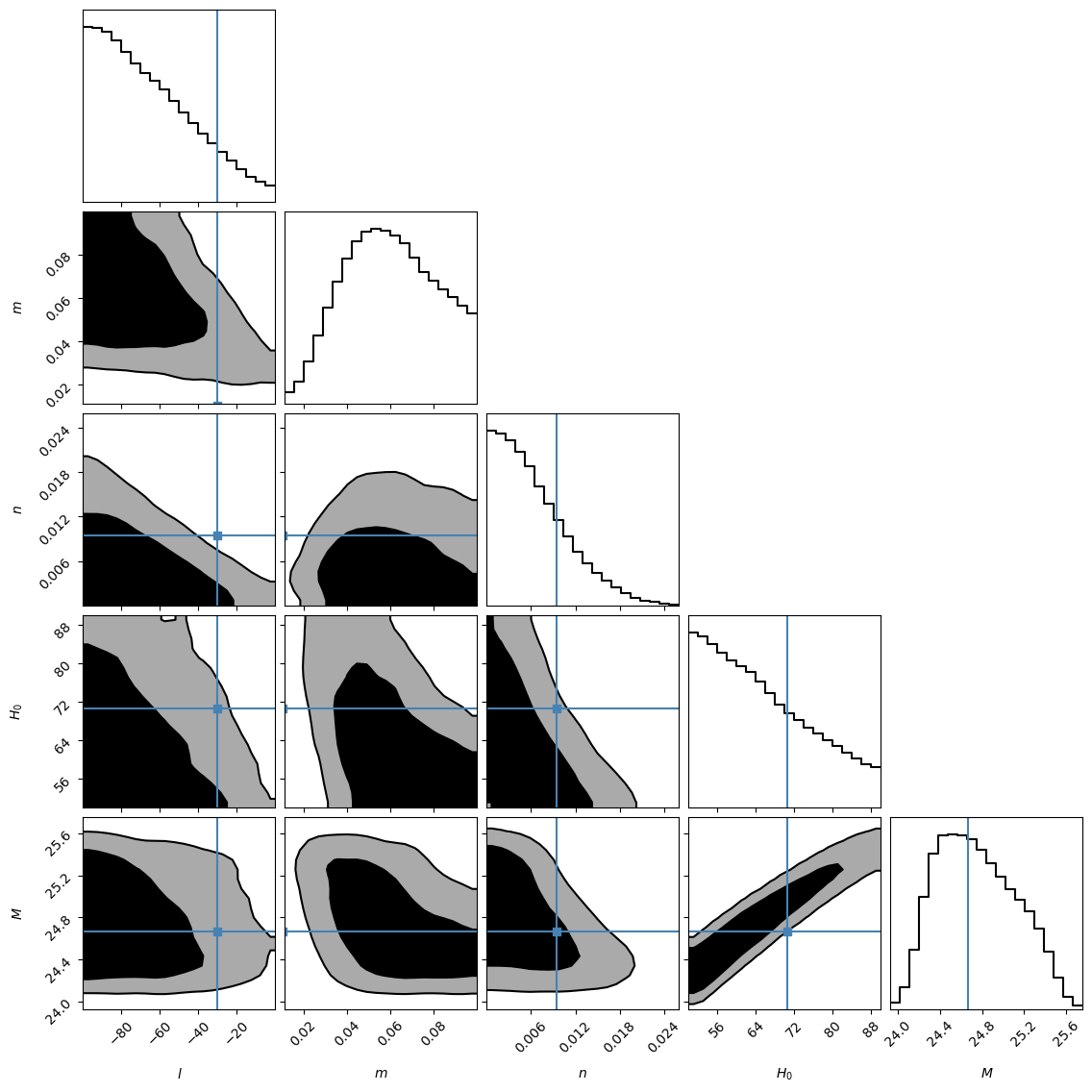}
	\caption{Posterior distributions and two–dimensional confidence contours for the model parameters obtained from the Union3 supernova data. The contours correspond to the $1\sigma$ and $2\sigma$ confidence levels.}
\end{figure}
\begin{figure}
	\centering
	\includegraphics[width=0.50\textwidth]{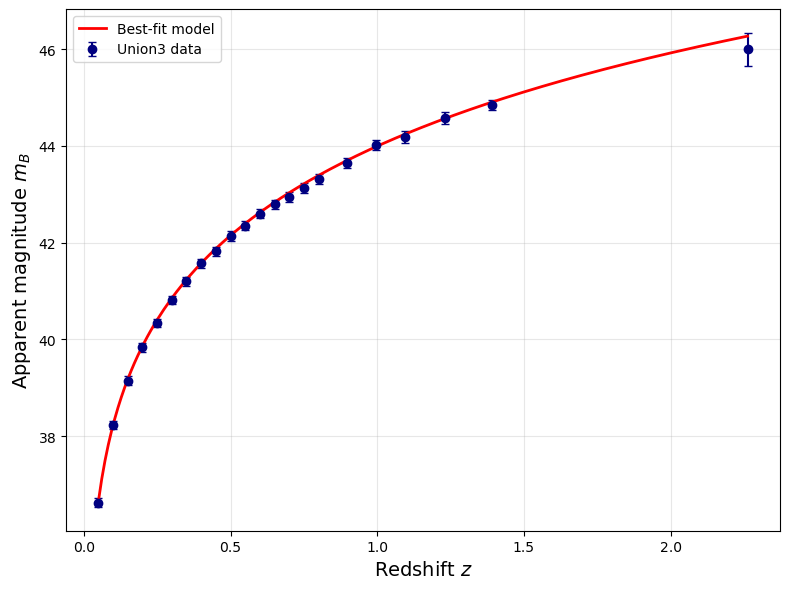}
\caption{Apparent magnitude $m_B$ of the Union3 binned Type Ia supernova sample plotted against redshift $z$. The blue points with error bars represent the observational data, while the red curve shows the theoretical prediction of the model using the best–fit parameters derived from the MCMC analysis.}
\end{figure}

Figure~4 and ~5 shows the Hubble diagram for the Union3 binned Type Ia supernova sample together with the best–fit theoretical prediction obtained from the MCMC analysis. The blue points with $1\sigma$ error bars represent the observed apparent magnitudes $m_B$ as a function of redshift, while the red curve corresponds to the model prediction using the best–fit parameters.

It is evident from the figure that the theoretical curve provides a good description of the observational data over the entire redshift range of the Union3 sample. The agreement between the model and the supernova observations indicates that the proposed cosmological framework can successfully reproduce the late–time expansion history of the Universe.

The uncertainties correspond to the $1\sigma$ confidence intervals derived from the marginalized posterior distributions. The results indicate that the supernova data alone provide relatively broad constraints on the cosmological parameters due to degeneracies among them. In particular, a strong degeneracy between the Hubble constant $H_0$ and the nuisance parameter $\mathcal{M}$ is observed, which is a well–known feature of supernova analyses.

To further tighten the constraints on the cosmological parameters and reduce degeneracies, it is necessary to combine the supernova observations with other cosmological probes such as baryon acoustic oscillations (BAO) and cosmic microwave background (CMB) measurements.
\section{Cosmological parameters}
\subsection{Deceleration Parameter}
The deceleration parameter depends on the higher derivatives of the scale factor $ a $ and it is defined as
\begin{equation}
    q(t) = -\frac{\ddot{a} a}{\dot{a}^2}= -1+\frac{\dot{H}}{H^2}
\end{equation}
In terms of red shift $ z $, the above Eq. implies
\begin{equation}
    q(z) = -1 + (1+z) \frac{H'(z)}{H(z)}
\end{equation}
\begin{figure}[H]
	\centering
	\includegraphics[width=8cm,height=6cm,angle=0]{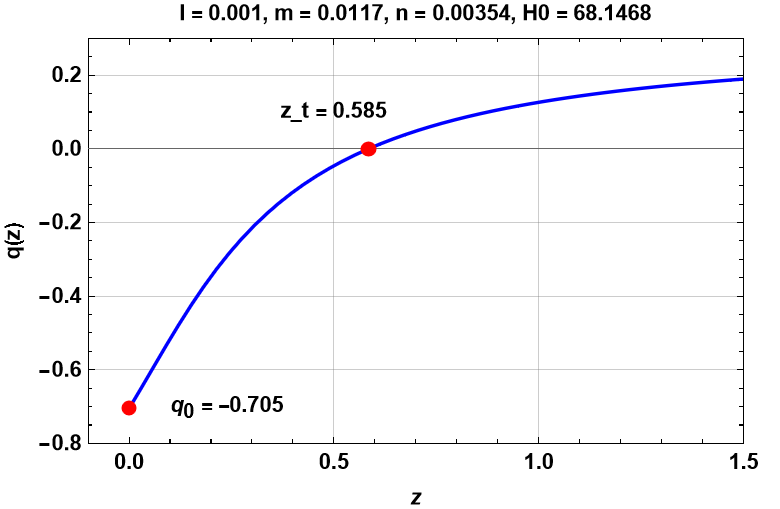}
	\caption{Deceleration parameter plot against redshift. Parameters estimation based on minimum $\chi^{2}$ for 46 Hubble data set.}
    \label{fig3}
\end{figure}
The figure (3) shows the deceleration parameter $q(z)$ plotted against redshift $z$. Transition Redshift ( $z_t = 0.585$ ) indicating the transition from a decelerated to an accelerated expansion of the universe. This is consistent with the current cosmological understanding where the universe transitioned to accelerated expansion at a redshift around z$\sim$0.5. The current deceleration parameter ($q_0 = -0.705$ ) is prominently displayed and reflects the current accelerated expansion (negative $q$). This value aligns well with the standard $\Lambda$CDM model predictions. The curve shows $q(z)$ increases monotonically with z, starting from $q_0$ at $z=0$, crossing q(z)=0 at $z_t$, and continuing toward positive values at higher redshifts. This behavior suggests that at higher redshifts, the deceleration parameter as positive, corresponding to a decelerated expansion dominated by matter.
\subsection{Density, Pressure and Equation of State parameters of the Model:}
The equation of the state parameter, defined as $\omega(z)= p(z) /\rho(z)$, characterizes the relationship between the pressure ($p$) and energy density $(\rho)$ of the cosmic fluid. This reflects the nature of the dominant energy component (e.g., matter, radiation, or dark energy). At higher redshifts, 
$\omega(z)$ approaches values typical of a matter-dominated universe $(\omega\sim0)$.
As the redshift decreases, $\omega(z)$ transitions toward negative values, indicating the dominance of dark energy with negative pressure driving the accelerated expansion.
The evolution of $\omega(z)$ highlights the transition from a decelerated (matter-dominated) to an accelerated (dark energy dominated) phase of the universe. Pressure $p(z)$ is derived from the bulk viscosity modified Einstein field equations and depends on the redshift-dependent Hubble parameter $H(z)$ and its derivative $H'(z)$. Negative pressure is essential for accelerated expansion of the universe. At higher the redshifts, $ p(z)$ starts near zero, corresponding to the negligible pressure of non-relativistic matter in the early universe.
As redshift decreases, $p(z)$ becomes increasingly negative, coinciding with the onset of dark energy dominance and accelerated expansion. The smooth evolution of $p(z)$ with decreasing redshift reflects the gradual emergence of negative pressure owing to bulk viscosity effects.\\
From Eqs. (\ref{8}) and (\ref{9}), the density $\rho$, pressure $p$, and the equation of state $\omega$ parameters of the universe as per our model are obtained  as follows:

\be
  \rho (z)=\rho_{c0}\frac{H(z)^2}{{H_0}^2};~~\rho_{c0} = \frac{3c^2H_0^2}{8 \pi G},
\ee
\be
p(z)= \rho_{c0} \frac{H(z)^2 \left(\frac{2 (z+1) H'(z)}{3 H(z)}-1\right)}{{H_0}^2},
\ee
and
\be
\omega (z) = \frac{p}{\rho} = \frac{2 (z+1) H'(z)}{3 H(z)}-1,
\ee
where $ \rho_{c0 }= \frac{3c^2H_0^2}{8 \pi G}$ is obtained as $8.73069*10^{-30} gm/cm^3$ for $H_0= 68.1468~ km/sec/Mpc.$\\

 The above expressions are equivalent reformulations of Eqs. (7) and (8), written in terms of the present critical density ( $\rho_{c0}$ ) for convenience in observational analysis. Although we adopt natural units ( $\frac{8\pi G}{c^4} = 1$ ) in the field equations, we reintroduce the critical density ($ \rho_{c0} $) in standard units to facilitate direct comparison with observational results.\\
 
We present the following figures for these parameters.
\begin{figure}[H]
	\centering
	a.\includegraphics[width=8cm,height=6cm,angle=0]{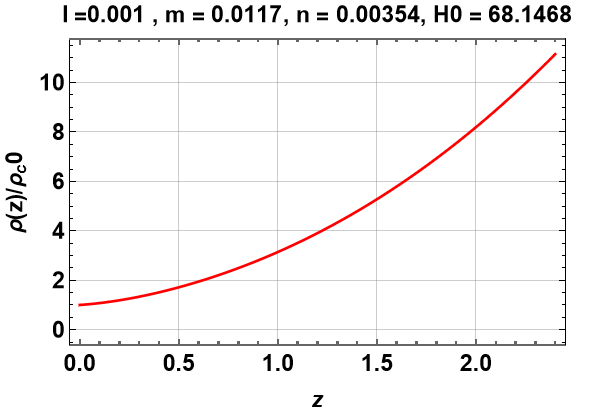}
    b.\includegraphics[width=8cm,height=6cm,angle=0]{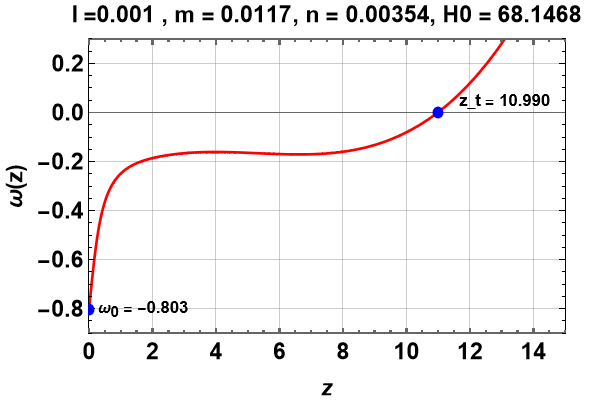}
    
\end{figure}
\begin{figure}[H]
	\centering
    c.\includegraphics[width=8cm,height=6cm,angle=0]{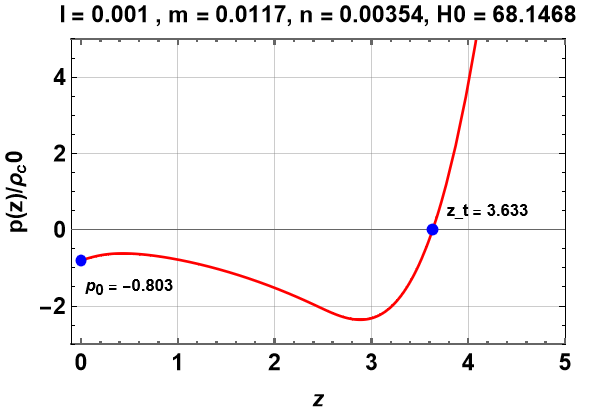}
	\caption{Density, Pressure and Equation of State parameters of the Model.}
	\label{fig4}
\end{figure}
The density plot in Fig (4a) illustrates the evolution of the cosmic energy density $\rho(z)$ as a function of the redshift $z$, providing insights into the distribution of energy components in the universe over time. The energy density $\rho(z)$ decreased monotonically as the redshift decreases.
At higher redshifts, $\rho(z)$ is significantly larger, reflecting the dominance of matter in the early universe when the scale factor $a(t)$ was small. As the universe expands, the density of matter dilutes and dark energy begins to dominate.
The plots Figs. (4b) and (4c) for the equation of state parameter $\omega(z)$ and pressure $p(z)$ provide valuable insights into the behavior of the cosmic fluid and its impact on the expansion of the universe.
The $\omega(z)$ and $p(z)$ plots provide complementary views of the same physical processes, that is, the transition from a matter-dominated phase to a dark energy-dominated phase. These plots demonstrate how bulk viscosity contributes to the negative pressure necessary for accelerated expansion. The behavior of $\omega(z)$ and $p(z)$ is consistent with the predictions of the modified cosmological models and aligns with observational constraints. These plots further validate the robustness of the bulk viscosity model in capturing the observed dynamics of the universe's expansion.
\section{State Finder analysis}
State finder diagnostics \cite{Sahni:2002fz, Alam:2003sc, Sahni:2008xx} provide a way to distinguish between different cosmological models by examining the parameters $r$ and $s$, which are derived from the Hubble parameter and its derivatives. These diagnostics are particularly useful for understanding the behavior of dark energy and its evolution  with cosmic time.
We define the following state finder parameter for our analysis \cite{Sahni:2002fz}:
\be
r(z)=q(z) (2 q(z)+1)+(z+1) \frac{\partial q(z)}{\partial z}
\ee
\be
s(z)=\frac{j(z)-1}{3 (q(z)-0.5)}
\ee
Now we plot the following:
\begin{figure}[H]
	\centering
	a.\includegraphics[width=8cm,height=8cm,angle=0]{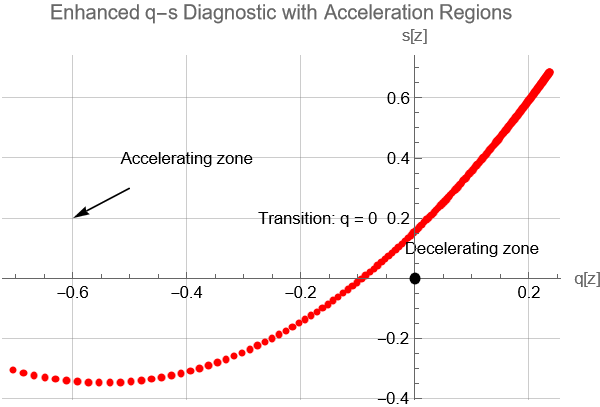}
    b.\includegraphics[width=8cm,height=8cm,angle=0]{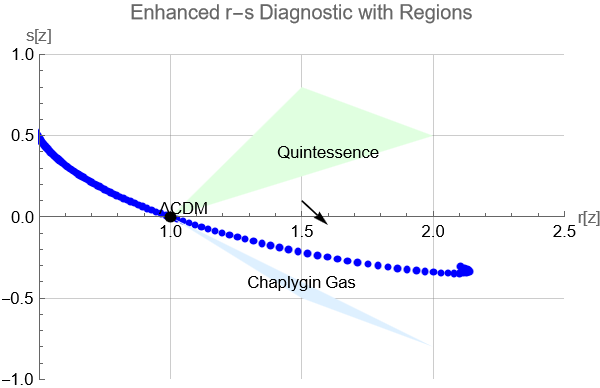}\\
    \label{fig5}
\end{figure}
\begin{figure}[H]
	\centering
	c.\includegraphics[width=8cm,height=8cm,angle=0]{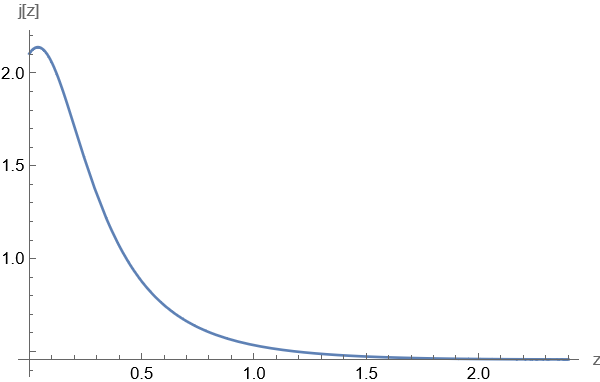}
	d.\includegraphics[width=8cm,height=8cm,angle=0]{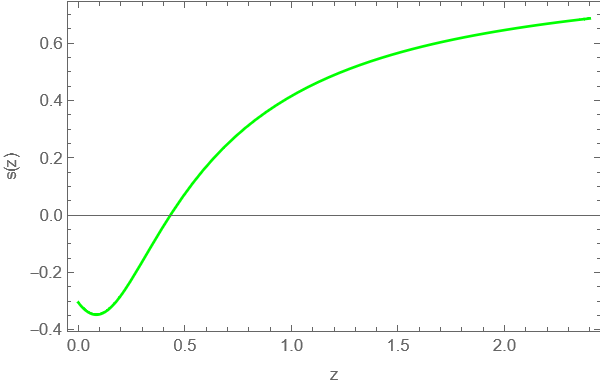}
	\caption{Plots for State finder analysis.}
	\label{fig5}
\end{figure}
The Fig (5a) shows the relationship between the deceleration parameter $q(z)$ and the state finder parameter $s(z)$. The transition from deceleration $(q>0)$ to  acceleration $(q<0)$ was clearly visible. On the right side of the transition point. $z_t$ is the deceleration zone$(q>0)$ whereas left side zone $(q<0)$ is the accelerating zone. The transition point $z_t$ where $q=0$ is explicitly marked, providing insight into the redshift at which the universe transitions from deceleration to acceleration. This diagnostic method does not assume a specific dark energy model, making it versatile for various cosmological scenarios. The separation of the acceleration and deceleration regions makes it easier to interpret the data and validate theoretical models.
The second plot in Fig (5b) shows the relationship between $r(z)$ and $s(z)$. Regions corresponding to different cosmological models are marked as:\\
$\Lambda$CDM: Located at $(r,s)=(1,0)$.\\
Quintessence: Region in the top right with $r>1, s>0$.\\
Chaplygin Gas: Region in the bottom left with $r>1, s<0$.\\
The plot helps to distinguish between different dark energy models ($\Lambda$CDM, quintessence, and Chaplygin gas). By plotting $r$ and $s$, the evolution of the cosmic acceleration was captured, allowing for comparison across models. The $\Lambda$CDM point provides a reference for testing deviations from the standard model.
 In our plot, the trajectory  passes near this point, which indicates that our model mimics $\Lambda$CDM at certain epochs. The $s<0 and r>1 $ typically imply a universe transitioning faster into an accelerated phase or with an earlier onset of dark energy domination. This points to non-standard cosmological models where dark energy evolves dynamically and significantly affects the cosmic expansion. These values often suggest models with phantom-like behavior ($\omega<-1$). This result highlights a departure from the$\Lambda$CDM paradigm.\\
 The jerk $j(z)$ plot in Fig (5c) starts near $j=1$ at low redshifts (for $\Lambda$CDM-like behavior) and deviates as 
$z$ increases, reflecting the model's deviation from the standard $\Lambda$CDM.For bulk-viscous models, 
$r(z)$ can exhibit behavior indicative of a phantom or quintessence-like evolution, with trajectories passing through or near $r=1$. At higher redshifts, $j(z)$ may exhibit larger deviations owing to the impact of the viscosity and matter-dominated dynamics.
At lower redshifts, $j(z)$ reflects the onset of acceleration with values approaching those expected in $\Lambda$CDM $(j=1)$. At higher redshifts, the plot captures the transition to deceleration, which is influenced by dominance of the matter .
Insights from Fig (5d) $s(z)$ and Fig (5c) $j(z)$ plots:Diagnostics of Cosmic Evolution:
These plots help distinguish between different cosmological models (e.g., $\Lambda$CDM, quintessence, and phantom models). \\
The trajectory on the $r-s$ plane depicts the entire evolution of the cosmological model \cite{Capozziello et al. 2011}. The fixed point $(1, 0)$ represents $\Lambda$CDM, but deviations from it and the route towards it represent the dynamics of dark energy or modified gravity. The trajectory's direction, curvature, and endpoint serve as powerful diagnostic tools to distinguish between rival cosmological hypotheses. The two models  have  identical present values of $H_{0}$ and $q_{0}$, but have distinct trajectories on the $r-s$ plane.

\section{Conversion of Redshift into Time and Age of the
		Universe:}
	We may get time in a billion years with the help of redshift as follows:\\
	The scale factor $a(z)$ is related to the red shift through the relation,
	$$1+z= \frac{a_0}{a(t)}$$
	differentiating this w.r.t time $`t'$ and using Hubble parameter $H=\frac{a'(t)}{a(t)}$, we get
	\be\no
	\frac{dz}{dt} = -(1+z)H(z)
	\ee
	
	This equation is integrated to yield the following relationship
	between elapsed time~$(t_0 - t_z$)~from the present and redshift
	$z$:
	
\begin{equation*}
t_{0}-t_{z}=\int_0^z \frac{1}{(z+1)H(z)} \, dz,
\end{equation*}
where $t_0$ is the present time at $z=0$ and $t_z$ is the time at
redshift $z$, $(t_0 -t_z$) is the elapsed time
from the present time at redshift $z$.
We note that the unit of the Hubble parameter $H(z)$ \small{ Km/sec/Mps}, which is a unit of reciprocal of time. We used the following
conversion formula to express the unit $\frac{Mps}{km/sec}$ of the reciprocal of the Hubble parameter $H(z)$ in time, that is, in billion years.
	
	$$\frac{Mps}{km/sec} = \small{978.462}~ \small{billion yrs}$$.

The current age of the universe  $t_{0}$ in Gyrs  is obtained  from the following: 
\begin{equation}\label{17}
t_{0}=\lim_{x \to \infty}\int_0^x \frac{978.462}{ (z+1)H(z)} \, dz
\end{equation}
By integrating Eq. (\ref{17}) and taking the limit, we obtain the the current age of the universe  as
  $ t_{0}$ =14.5734 for $l=0.001$, $m=0.0117$, $n=0.0035$, and  $H_0=68.148$.\\
  We present the following figure to display our findings graphically.
   \begin{figure}[H]
	\centering
	\includegraphics[width=8cm,height=6cm,angle=0]{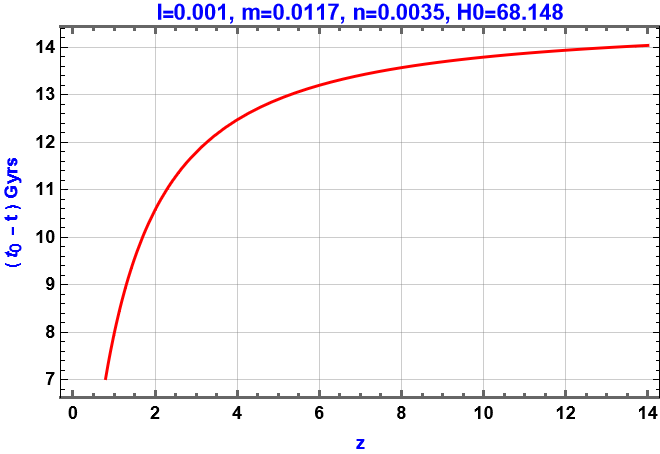}
    
	\caption{Plot for redshift $z$ versus time in gigayears.}
    \label{fig2}
\end{figure}
Figure 6 shows the variation of $  (t_{0}-t) $ in Gyrs over redshift $ z $ . The plot approaches $ t_{0}$ = 14.5734 described earlier. According to WMAP data, the empirical value of the current age of the universe is $ t_{0}=13.73^{+.13}_{-.17} $ $ Gyrs $.
\section{Conclusion:} The study explored the role of bulk viscosity in a dust-filled universe using the FLRW metric. The following is a concise summary: 
\begin{itemize}
\item We begin with two objectives: First we investigated the influence of bulk viscosity ($\tilde{p}= -3(l + m (H'(t) + H^2) + n H^2) H$) on on cosmic expansion: second, we determined unknown parameters through estimations using observational data sets.\\
\item The given bulk viscosity model is well-structured to explore late-time cosmic acceleration and alternative dark energy effects. The dependence on both $H$ and $H'$
makes it sufficiently flexible to capture different evolutionary phases of the universe. The choice of l,m,n determines whether the model behaves more like 
$\Lambda$ CDM, a dynamical dark energy model, or a modified gravity scenario.
\item We formulated Einstein field equations using bulk viscosity and have solved them numerically in terms of the redshift $z$.\\
\item We used two types of data sets for parameter estimations. 46 Hubble data points for $H(z)$ to estimate parameters via $\chi^{2}$ minimization and refined with Markov chain Monte Carlo (MCMC) simulations and 1701 pantheon+ supernova data points for distance modulus $\mu(z)$, analyzed similarly.\\
\item The transition redshift ($z_t = 0.585$) for the deceleration parameter  q(z) where it crosses zero, indicates a shift from deceleration to acceleration. The present value ($q_0=-0.705$) aligns with $\Lambda$ CDM predictions.\\
\item We derived expressions for the density ($\rho$), pressure ($p$), and  equation of state ($\omega$). The $\omega$(z) and $p(z)$  provide complementary views of the same physical processes: the transition from a matter-dominated phase to a dark energy-dominated phase. They demonstrated the contribution of bulk viscosity to the negative pressure necessary for the accelerated expansion. \\
\item We carried out  state finder diagnostics for our model and explored parameters $r$ and $s$ to differentiate between cosmological models such as  $\Lambda$CDM and Quintessence.
\item  Additionally, we estimated the present Hubble constant, $H_0$ , to be 68 km/s/Mpc based on the Hubble data set and approximately 73 km/s/Mpc using the Pantheon+ datasets. This disparity highlights the ongoing Hubble tension—a discrepancy between locally measured values of  $H_0$(e.g., via supernovae) and values inferred from the early universe (e.g., CMB observations). Our results align with the broader trend of higher  $H_0$ values from late-universe data compared with early-universe predictions, underscoring the need for further investigation into the underlying physical or systematic causes of this tension.
\item The current age of the universe obtained  from  our model was $t_{0}$ = 14.5734 Gyrs. According to WMAP data, the empirical value of the current age of the universe was $ t_{0}=13.73^{+.13}_{-.17} $ $ Gyrs $.   Precision measurements of the cosmic microwave background (CMB), along with baryon acoustic oscillations and supernova data, indicate a current cosmic age of around 13.8 Gyr under the mainstream $\Lambda$CDM paradigm. However, this estimate is essentially model-dependent and is based on the assumptions of large-scale homogeneity, a cosmic constant, and general relativity as the right explanation of gravity at all scales.
Some study implies the oldest globular clusters and individual stars may imply ages $\geq $14 Gyr, potentially exceeding $\Lambda$CDM's $\sim$13.8 Gyr unless uncertainties are substantial or physics is altered. See some references \cite{R10A,R10B,R10C,R10D}.   
\item With the increasing precision of cosmological observations, several tensions in the key cosmological parameters have emerged. One of the most significant aspect is  Hubble tension. This tension refers to the inconsistency between the values of the Hubble constant $H_{0}$ inferred from early-universe observations and those obtained from local, late-time measurements. Despite substantial improvements in observational accuracy, these discrepancies persisted and may indicate new physics beyond the standard cosmological model. A Planck Collaboration's analysis of CMB data within the $\Lambda$CDM framework yields $H_{0} = (67.4 \pm 0.5) km s^{-1} MPc^{-1}$ at 68\% confidence level (CL) \cite{NA:2020}. In contrast, the SH0ES Collaboration (Supernovae $H_{0}$ for the Equation of State ) found a significantly higher value of $H_{0} = (73.04 \pm 1.04) km s^{-1} MPc^{-1}$ at 68\% CL, using the three-rung distance ladder method with Cepheids \cite{R10}. A more comprehensive review addressing Hubble tension can be found in \cite{R11,R12,R13,R14,R15}. 
Future large-scale surveys, such as Euclid and next-generation LSS investigations, will offer precise measurements of the expansion history and the growth of cosmic structures. These data can be used to recover higher-order kinematic parameters, such as the statefinder pair (r,s), and compare the anticipated trajectories of the current model against the $\Lambda$ CDM fixed point. Furthermore, redshift-space distortion and weak-lensing observations break the degeneracies between background expansion and growth, providing a rigorous observational test of the concept. For a detailed study, we refer the readers to some reference \cite{R16,R17,R18}.

\end{itemize}	

Finally we state that the bulk viscosity model provides a viable framework for understanding cosmic acceleration and fits the observational data well. Transition redshift and parameter estimations align with the established cosmological understanding, indicating consistency with $\Lambda$CDM while exploring deviations.

\section*{Declaration of conflict of interest}
The authors declare that they have no known competing financial interests or personal relationships that could have influenced the work reported in this study.

\section*{Acknowledgments}
The authors are grateful to the IUCCA in Pune, India, for providing facilities during a visit where a portion of the study was completed. The authors respectfully acknowledge gratitude to the reviewers for their insightful remarks that improved the manuscript in its current form. 


\end{document}